\documentclass[aps, superscriptaddress, showpacs, showkeys, twocolumn]{revtex4}

\usepackage{graphicx}

\begin{document}

\title{Structure and giant magnetoresistance of granular Co-Cu nanolayers prepared by cross-beam PLD}

\author{A.~Jesche}
\email{jesche@cpfs.mpg.de}
\altaffiliation[present address: ]{Max-Planck-Institut f\"ur Chemische Physik fester Stoffe, D-01187 Dresden, Germany}
\affiliation{Institut f\"ur Strukturphysik, Technische Universit\"at Dresden, D-01062 Dresden, Germany}
\author{A.~Gorbunoff}
\affiliation{Hochschule f\"ur Technik und Wirtschaft Dresden, D-01069 Dresden, Germany}
\author{A.~Mensch}
\affiliation{Institut f\"ur Werkstoffwissenschaft, Technische Universit\"at Dresden, D-01062 Dresden, Germany}
\author{H.~St\"ocker}
\author{A. A.~Levin}
\author{D. C.~Meyer}
\affiliation{Institut f\"ur Strukturphysik, Technische Universit\"at Dresden, D-01062 Dresden, Germany}

\begin{abstract}
A series of Co$_{x}$Cu$_{100-x}$ ($x$ = 0, 40...75, 100) layers with thicknesses in-between 13\,nm and 55\,nm were prepared on silicon substrates using cross-beam pulsed laser deposition. Wide-angle X-ray diffraction (WAXRD), transmission electron microscopy (TEM) and electrical transport measurements revealed a structure consisting of decomposed cobalt and copper grains with grain sizes of about 10\,nm. The influence of cobalt content and layer thickness on the grain size is discussed. Electron diffraction (ED) indicates the presence of an intermetallic Co-Cu phase of Cu$_3$Au structure-type. Thermal treatment at temperatures between 525\,K and 750\,K results in the progressive decomposition of Co and Cu, with an increase of the grain sizes up to about 100\,nm. This is tunable by controlling the temperature and duration of the anneal, and is directly observable in WAXRD patterns and TEM images. A careful analysis of grain size and the coherence length of the radiation used allows for an accurate interpretation of the X-ray diffraction patterns, by taking into account coherent and non-coherent scattering. The alloy films show a giant magnetoresistance of 1...2.3\,\% with the maximum obtained after annealing at around 725\,K. 
\end{abstract}



\maketitle

\section{Introduction}
With the discovery of the giant magnetoresistance (GMR) effect in multi- \cite{gruenberg86} and single-layer \cite{Xiao92} systems, thin films of cobalt and copper have been investigated for basic research and applications \cite{chen2008}. 
The value of GMR, which varies between 1\,\% and 20\,\% in granular layers of Co and Cu, depends strongly on the preparation and post-process annealing.
In addition to the electronic properties, these layers are also interesting from the structural point of view. Despite being isomorphic metals with quite similar lattice parameters (cobalt: space group $Fm\bar3m$ , $a$ = 3.5446\,\AA; copper: $Fm\bar3m$, $a$ = 3.6148\,\AA\,\cite{pearson}) cobalt and copper are immiscible as bulk materials \cite{massalski}. However, it is well known that thin layers can behave differently, especially when energetic deposition methods like pulsed laser deposition (PLD) are used. Thus, in addition to the decomposed Co and Cu grains \cite{Garcia02} found in thin layers, a solid solution (mixed crystal) \cite{pohor98} and evidence for an intermetallic alloy \cite{yan07} have been reported for Co-Cu systems. The use of cross-beam PLD (CB-PLD) for the preparation of alloy Fe-Cr nanolayers has led to the discovery of crystalline phases not reported before \cite{gorbunov02,levin02}. Therefore, the aim of the present work was to show, whether the formation of metastable phases by using CB-PLD is also observable for Co-Cu layers. 

We present a growth procedure, determine the lattice spacing and grain size, and discuss the effect of a post-process annealing on structure and magnetoresistance of Co-Cu nanolayers with different compositions.

\section{Experimental Procedure}

\subsection{Sample preparation}
The Co-Cu alloy films were prepared by CB-PLD. A detailed description of this technique can be found in Ref.\,\cite{gorbunoffhabil}. 
CB-PLD allows the simultaneous and droplet-free deposition of two materials, in our case Co and Cu.
In this technique, after ablation from two separated targets, the materials are mixed to variable proportions in the colliding laser plumes, thereby resulting in a mixed Co-Cu plume that is directed towards the substrate. 
Thermally oxidized (001) silicon wafers with a thickness of the amorphous silicon oxide layer of 500\,nm were used as substrates. 
All depositions were performed at room temperature (RT) in a high vacuum (pressure $p < 10^{-6}$\,mbar). 
A Nd:YAG laser with a wavelength of $\lambda = 1064$\,nm and an intensity of $\approx1$\,GW/cm$^{2}$ at the target was used for the ablation. The repetition rate of the laser was 10 shots per second yielding an average film-deposition rate of $\approx 0.3$\,nm/min on the substrate.
In order to allow for uniform ablation, the target was rotated with a speed of $\approx 0.5$ rotations per second combined with a simultaneous periodic translation with a rate of 0.1\,mm/s. 
The substrate was positioned at a distance of 10\,cm from the target and 3\,cm away from the origin of the resulting laser plume. In order to get a more homogeneous film, the substrate was rotated with a speed of $\approx 1$\, rotation per second.
By varying the ablation geometry, compositions of 40\,\% to 75\,\% cobalt could be realized. 
A preparation of Co-Cu alloy films containing more Cu is difficult because of the enhanced resputtering of Cu compared to Co.
In addition, pure Co and Cu films were prepared.

A post-process anneal was carried out in a tubular furnace at temperatures from $T = 525$\,K up to $T = 725$K in steps of 25\,K under high vacuum conditions ($p \approx 10^{-6}$\,mbar). After an anneal duration of 75\,min, the samples were cooled down with an average rate of 10\,K/min. A final thermal treatment was carried out at 750\,K for 16\,h in order to reach a stable state. 

\subsection{Characterization}
The thickness of the deposited layers was determined by means of X-ray reflectometry to vary between 13\,nm and 55\,nm and did not change significantly after annealing.
The lattice spacing and crystallite size of the Co-Cu films were determined by wide-angle X-ray diffraction (WAXRD) using a D8-Advance diffractometer (Bruker AXS) with Cu-$K\alpha$ radiation in divergent beam geometry. The Si\,004 reflection was used as an internal standard for correcting $2\theta$ angle reflection positions. A coupled (symmetric) $\omega$-$2\theta$ scan mode was used, where $\omega$ is the angle of the incident X-ray beam with respect to the surface, 2$\theta$ is the angle between incident and reflected X-ray beam and $\omega$=$\theta$ for the symmetric scan mode. Therefore, only lattice planes oriented parallel to the sample surface were taken into account in this scan mode.
Asymmetric $2\theta$ scan modes ($\omega$ is fixed at a value corresponding to the maximum intensity of the respective rocking curve measured) did not show any influence on the X-ray reflection positions observed in the corresponding coupled scans.
No additional WAXRD reflections were found by asymmetric scans.

The chemical composition of the layers was determined by means of energy dispersive X-ray spectroscopy (EDX) using a semiconductor detector (Noran Voyager 2000). 
Transmission electron microscopy (TEM) and electron diffraction (ED) measurements were made in a Zeiss EM 912 OMEGA microscope in order to determine structural parameters, homogeneity and grain size of the films. As transmission of the silicon substrate with the electron beam, or the removal of the deposited layer from the substrate, are impossible, a NaCl crystal was coated together with the substrate. After deposition of the Co-Cu layer, the NaCl crystal was dissolved in water and the Co-Cu layer was put onto a TEM mesh.

The electric transport measurements were performed at temperatures from 10\,K to 300\,K in magnetic fields up to $\mu_0H = 12$\,T by using a VSM-Maglab cryostat (Oxford Instruments). The resistivity was measured by a standard 4-point contact method using conductive silver paint. In order to prevent silver contamination, some of the measurements were carried out using a simple 2-point contact method. A comparison of both techniques did not show significant differences.
The characterization was performed $ex\,situ$ because phases which are (meta)stable at ambient conditions were the focus of our investigations. Structure and composition did not change at ambient conditions over a period of one year. 

\section{Results - as-prepared samples}

\subsection{Structural parameters obtained by WAXRD}

\begin{figure}
  \includegraphics[width=7.5cm]{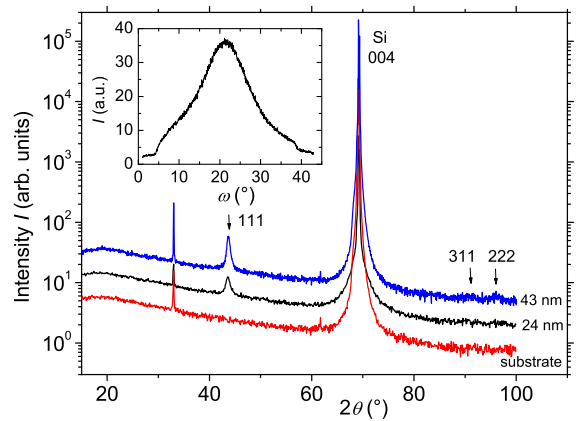}
    \caption{\label{fig1_diffr_typ} (Color online) WAXRD patterns of as-prepared Co-Cu films with different thickness $t$ and composition (Co$_{59}$Cu$_{31}$, $t = 24$\,nm and Co$_{40}$Cu$_{60}$, $t = 43$\,nm). For comparison, the WAXRD pattern of the Si substrate is shown (bottom). The reflection indices of the film reflections observed are indicated. The inset shows a typical rocking curve of the 111 reflection of a Co$_{40}$Cu$_{60}$ film.}
\end{figure}

Typical WAXRD patterns of two characteristic as-prepared Co-Cu nanolayers with different thickness and composition are shown in Fig.\,\ref{fig1_diffr_typ}. 
Note, in contrast to the known X-ray investigations on thin Co-Cu films from literature, our layers are rather thin. 
Due to the preferred orientation (fibre texture) and the small scattering volume of thin films, only a limited number of reflections are observable in the WAXRD patterns. Nevertheless, the WAXRD patterns show reflections not related to the substrate.
For all Co$_x$Cu$_{100-x}$ compositions investigated, the non-substrate reflection with the highest intensity can be indexed as the 111 reflection of the face-centred cubic (\textit{fcc}) Co-Cu lattice.
The inset in Fig.\,\ref{fig1_diffr_typ} shows a rocking-curve around this reflection, revealing on the one hand that the texture is strong enough to inhibit the detection of other WAXRD reflections. On the other hand, the rocking curve is broad in comparison to an epitaxially grown layer or a single crystal and, therefore, the intensity of the scattered radiation (for a given $\omega$-2$\theta$ coupling) is low. 
These facts are responsible for the difficulties associated with using XRD for thin layers. Hence, a careful optimization of step width, sampling time and slits is necessary in order to obtain adequate results.

For layers with a thickness of less than 30\,nm, only the 111 reflection of the face-centred cubic (\textit{fcc}) lattice is observable (see second curve in Fig.\,\ref{fig1_diffr_typ} as an example). Nevertheless, the crystallographic structure of the layers is assumed to be similar to the structure of bulk \textit{fcc} Cu and Co. This approach is supported by the fact that the observed (111) lattice spacing of pure Co and pure Cu layers is close to the values of the corresponding bulk materials. 
For the pure Co sample, the observed reflection can correspond either to the 111 reflection of the Co \textit{fcc} structure or to the 0002 reflection of the Co hexagonal structure. Both structures are known to exist in thin cobalt layers \cite{lamelas89,meny92,fauster93}. 
The upper pattern in Fig.\,\ref{fig1_diffr_typ} is a typical WAXRD pattern of a layer with a thickness of more than 40\,nm in which weak 311 and 222 Co-Cu reflections are observed. Their low signal-to-noise ratio does not allow a quantitative analysis of the structural parameters. 
The lower pattern in Fig.\,\ref{fig1_diffr_typ} belongs to an uncoated silicon substrate. This shows that the substrate reflections are up to three orders of magnitude higher than those belonging to the Co-Cu nanolayers.

\begin{figure}
  \includegraphics[width=7.5cm]{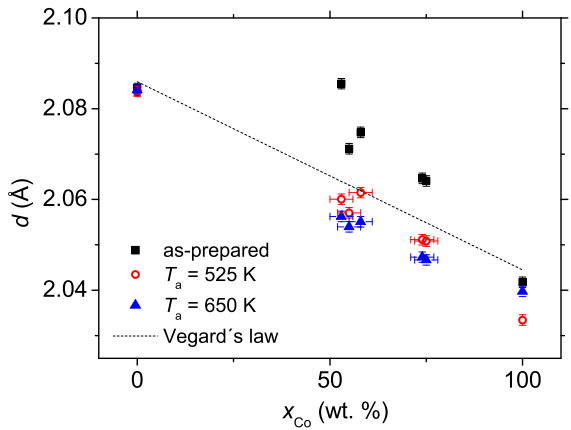}
    \caption{\label{fig2} (Color online) Lattice spacing of the (111) atomic planes of Co$_{x}$Cu$_{100-x}$ layers, as-prepared and after annealing at temperatures indicated (duration of 75 min). In the case of the pure Co film, the lattice spacing corresponds either to the (111) atomic plane of the Co \textit{fcc} structure or to the (0002) atomic plane of the Co hexagonal structure (see text). The estimated standard deviations (e.s.d.) of the experimental points are given by the horizontal and vertical bars. For comparison, the lattice spacings calculated according to Vegard's law are shown by the dotted line. For the calculations, the unit cell parameters of bulk \textit{fcc} Cu and Co were used \cite{pearson}.}
\end{figure}

Instead of the individual 111 Cu and 111 Co reflections that would be expected due to the immiscibility of the deposited materials, one 111 \textit{fcc} reflection is clearly observed in the WAXRD patterns of the Co-Cu layers. In general, this could be caused by the formation of a metastable Co-Cu solid solution or by size effects of the scattering crystallites. 
The determined (111) lattice spacing obtained for different Co$_{x}$Cu$_{100-x}$ compositions is shown in Fig.\,\ref{fig2} together with Vegard's law calculated for a hypothetic Co-Cu mixed crystal (see discussion). The strong deviation for the  Co$_{51}$Cu$_{49}$ composition (as-prepared) from Vegard's law, where the experimental lattice spacing is larger than that of pure copper, renders the existence of a solid solution very unlikely. 

It should be noted, that any strain present in the crystallites in the film could contribute to the deviation observed. By assuming the absence of strain in films after annealing at temperatures higher than 525\,K (see Fig.\,\ref{fig2}), the strain in as-prepared films was estimated from the lattice parameter change to be 0.5...1.5\,\%. An attempt to estimate the crystallite size in as-prepared films, based on these strain values, led to an unrealistic model of infinite crystallites (i.e. the strain was overestimated). Thus, the strain is significantly smaller than 0.5\% and could not be evaluated accurately using the data obtained.

\begin{figure}
  \includegraphics[width=7.5cm]{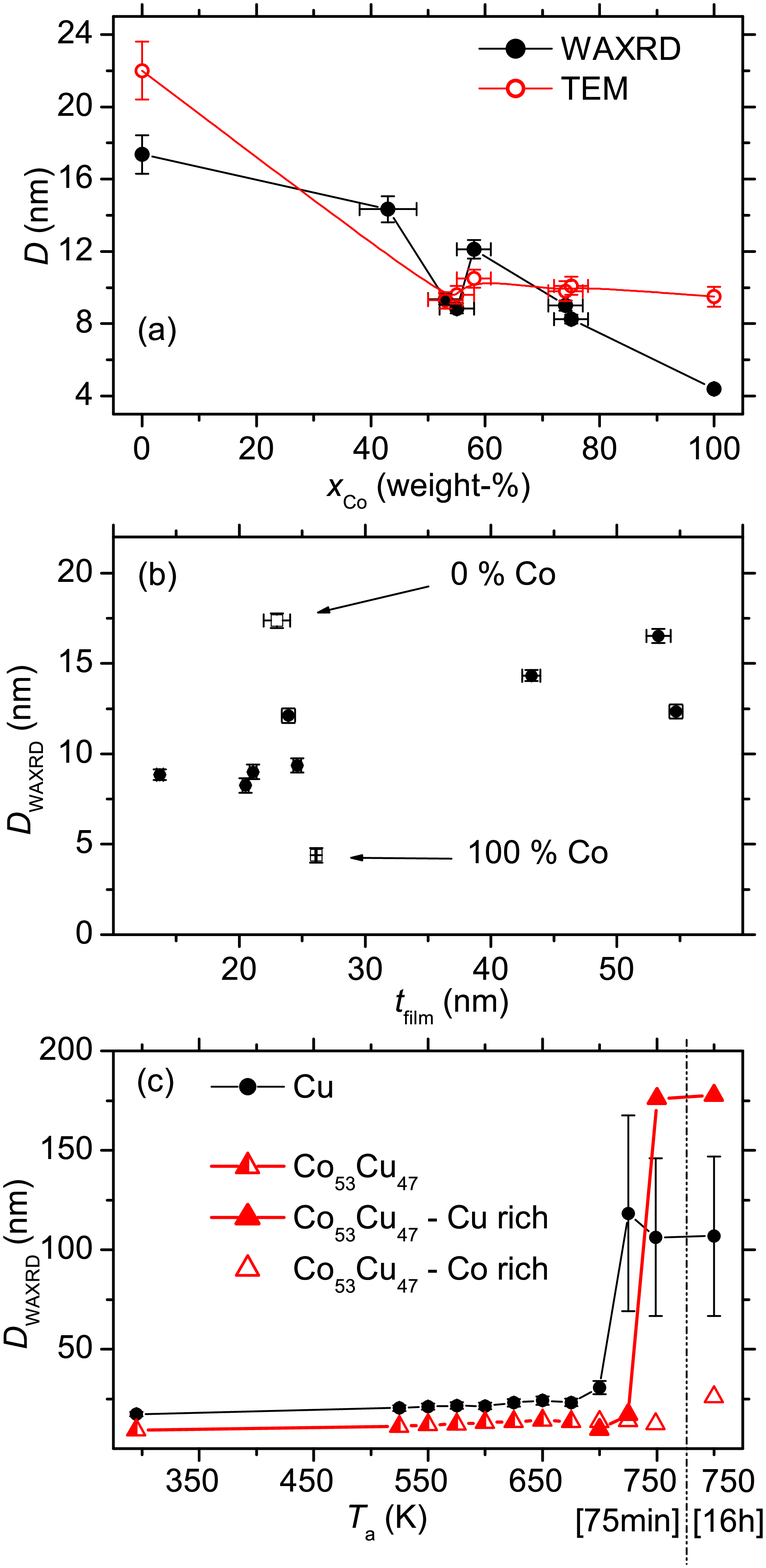}
    \caption{\label{fig3} (Color online) 
(a) Comparison of the crystallite sizes $D$ calculated by the Scherrer formula from WAXRD data and grain sizes measured by TEM for as-prepared Co$_{x}$Cu$_{100-x}$ films. 
(b) Crystallite size in as-prepared films as a function of the layer thickness.
(c) Development of the crystallite size after annealing at temperatures $T_{a}$ for Co$_{53}$Cu$_{47}$ and pure Cu layers (see Section 4.1). The lines in (a) and (c) are guides to the eye. The E.s.d.'s of the experimental points are shown by horizontal and vertical bars.}
\end{figure}

If strain is neglected, the crystallite size $D$ for Co$_{x}$Cu$_{100-x}$ films can be estimated by using the Scherrer equation
\begin{equation}
 D = 0.9\lambda/(B\cos\theta),
\end{equation}
where $B$ is the full width at half maximum (FWHM) of the WAXRD reflection. In order to take into account the instrumental broadening caused by the experimental setup, the observed FWHM ($F\!W\!H\!M_\mathrm{obs}$) of the WAXRD reflection was corrected by using the FWHM of the silicon 004 substrate reflection ($F\!W\!H\!M_\mathrm{instr}$). Since for all reflections $F\!W\!H\!M_\mathrm{obs}/\beta < 0.64$ ($\beta$ is the integral breadth of the reflection), the reflection profiles could be described as Lorentzians \cite{langford}. Following the procedure described in \cite{warren} for X-ray reflections of Lorentzian profile shape, the reflection FWHM was corrected by 
\begin{equation}
 B = F\!W\!H\!M_\mathrm{obs}\,-\,F\!W\!H\!M_\mathrm{instr}.
\end{equation}

The crystallite sizes ($D_\mathrm{WAXRD}$) obtained for different compositions, together with the grain size obtained by TEM ($D_\mathrm{TEM}$, see next paragraph), are shown in Fig.\,\ref{fig3}(a). The crystallite/grain size does not exceed the thickness of the layers.
We measured the thickness of the sample containing 53\,\%\,Co to be 25\,nm, whereas the thickness of the sample containing 55\,\% Co is only 13\,nm. Nevertheless, their crystallite sizes are very similar (Fig.\,\ref{fig3}(a)), which shows that the crystallite size is more strongly affected by the Co content than by the layer thickness. 
However, there is a weak correlation between layer thickness and crystallite size observable (Fig.\,\ref{fig3}(b)). 
Within the accuracy of our measurements the crystallite/grain sizes of the alloy systems lie between the grain sizes of the pure compounds.

\subsection{Transmission electron microscopy}

\begin{figure*}
  \includegraphics[width=\textwidth]{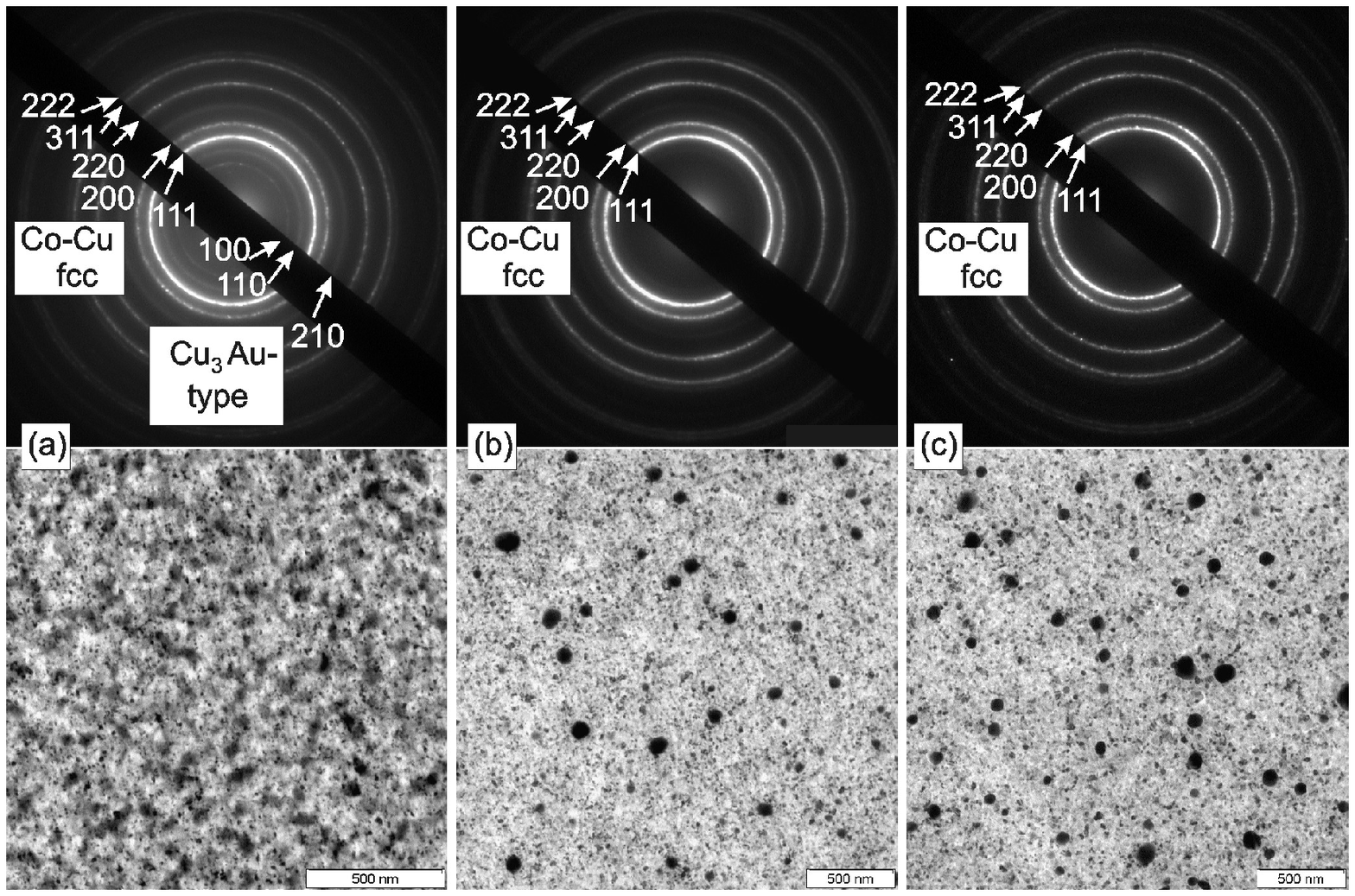}
    \caption{\label{fig4} TEM bright field images (lower part) and ED patterns (upper part) of a Co$_{59}$Cu$_{41}$ layer (a) as-prepared, (b) after thermal treatment for 15 min at 725\,K and (c) after thermal treatment at 875\,K. The diffraction rings not belonging to the \textit{fcc} lattice disappear after thermal treatment at 725\,K.}
\end{figure*}

Another approach used to measure the crystal grain sizes was TEM. The dark grains in the TEM micrographs (Fig.\,\ref{fig4}) fulfill the Bragg condition, i.e. the contrast is not induced by absorption differences. The grain size was determined by averaging the size of 20 randomly chosen grains. These are shown in Fig.\,\ref{fig3}(a) as a function of cobalt content.
It should be noted that the grain size determined by TEM is in most cases not the same quantity as the crystallite size obtained by WAXRD. The grains can contain more than one crystallite. As can be seen in Fig.\,\ref{fig3}(a), for all alloy compositions, the size of crystallites and grains determined by WAXRD and TEM, respectively, are equal in the limits of the estimated standard deviations (e.s.d.). This highlights the equivalence of the crystallite and grain sizes for the deposited alloy Co$_{x}$Cu$_{100-x}$ nanolayers. At the same time, the grain sizes of the pure Co and Cu films are about two to three times larger than the crystallite sizes, indicating the formation of large grains containing 2-3 crystallites.

ED-patterns (Fig.\,\ref{fig4}(a)) show the Bragg reflections which can be well indexed using an \textit{fcc} lattice in agreement with the WAXRD patterns.  Three additional reflections violating the extinction law of the fcc lattice were found for all compositions (as-prepared films). The assumption of an additional intermetallic phase of Cu$_3$Au-type and unit cell parameter $a = 3.55$\,\AA, theoretically predicted by Yan $et$ $al.$ \cite{yan07}, is in good agreement with the experimental data. As the Cu$_3$Au structure-type is a superstructure of the \textit{fcc} lattice, the similar atomic scattering factors of Co and Cu cause a weak intensity of the 100, 110 and 210 Bragg reflections which cannot be resolved by WAXRD. However, the different substrate (NaCl) also could be responsible for the occurrence of these phases compared to the case of the WAXRD investigated layers on amorphous silicon oxide substrates.  

\subsection{Electrical transport measurements -- giant magnetoresistance}

\begin{figure}
  \includegraphics[width=7.5cm]{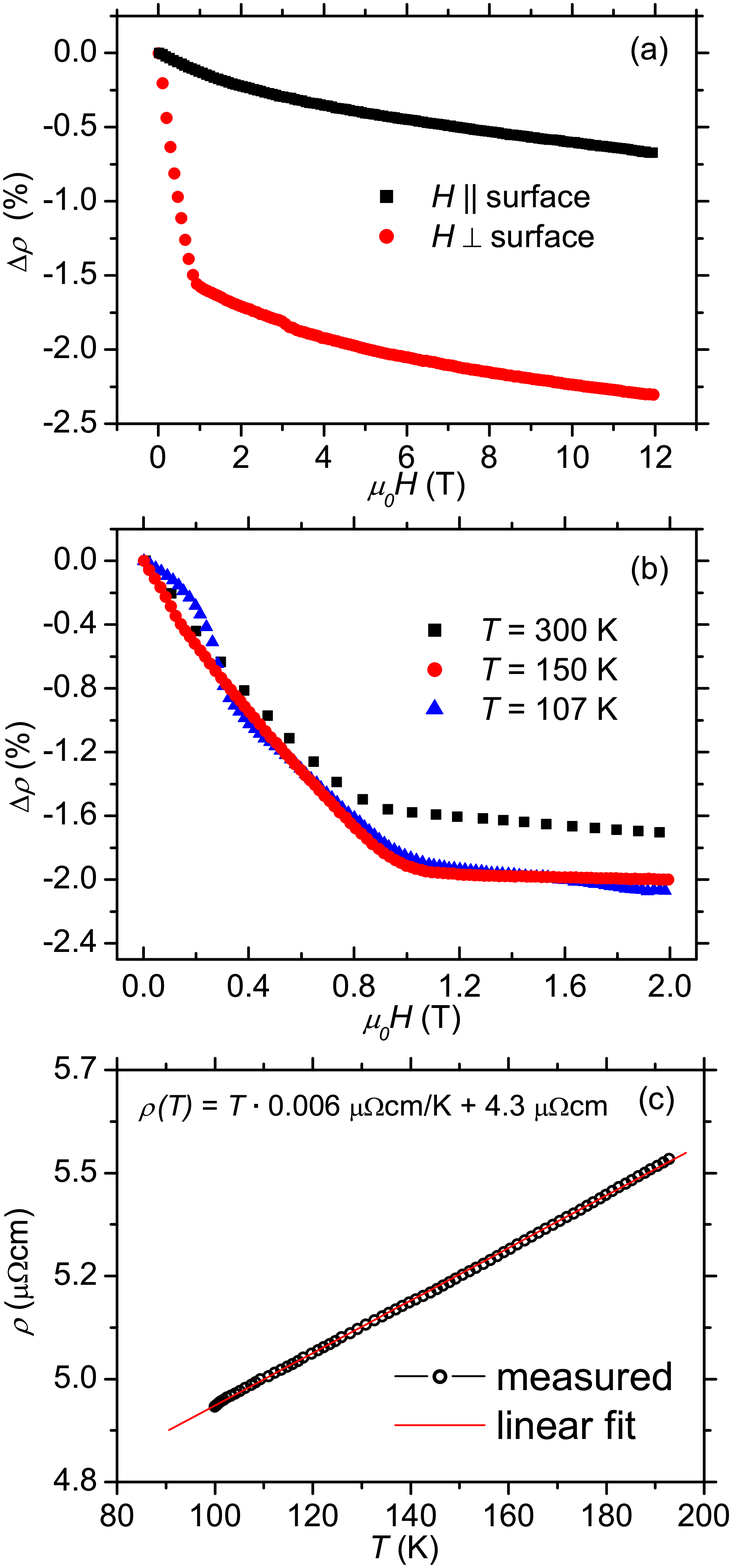}
    \caption{\label{fig5} (Color online) Nanolayer Co$_{59}$Cu$_{41}$. (a) Change of the electrical resistivity $\Delta\rho$ at RT for magnetic fields perpendicular and parallel to the surface. (b) Change in the electrical resistivity as a function of applied magnetic field for three different temperatures. The magnetic field was oriented perpendicular to the sample surface. (c) The electrical resistivity as a function of temperature in zero magnetic field.}
\end{figure}

The electrical transport measurements are shown in Fig.\,\ref{fig5} for nanolayered Co$_{59}$Cu$_{41}$ (as-prepared).
The development of the electrical resistance at RT as a function of applied magnetic field is shown in  Fig.\,\ref{fig5}(a) for two different orientations of the field. The change in the resistivity $\Delta\rho$ is defined as 
\begin{equation}
\Delta\rho = \frac{\rho(B)-\rho(B=0)}{\rho(B=0)}. 
\end{equation}
For a magnetic field oriented perpendicular to the surface and thus perpendicular to the electrical current, the resistivity rapidly decreases by $\Delta\rho = -1.6$\,\% at $\mu_0H = 0.9$\,T followed by a moderate decrease to $\Delta\rho = -2.3$\,\% at $\mu_0H = 12$\,T. This decrease is most likely caused by the GMR effect, because the anisotropic magnetoresistance (AMR) exhibits a smaller decrease in the resistivity and smaller saturation fields. Literature values for a pure Co layer prepared by PLD are $\Delta\rho = 0.5$\,\%, with a saturation field of $\mu_0H = 0.03$\,T \cite{krieger04}. The stronger GMR for a magnetic field aligned perpendicular to the surface is a thin film phenomenon and/or caused by the texture (the granular GMR should be isotropic). Even for high fields $\mu_0H \approx 12$\,T no saturation is observed, which is typical for the GMR in granular layers. 

For the following measurements, the magnetic field was applied perpendicular to the surface. Magnetic field dependencies of the GMR effect at RT and lower temperatures are shown in Fig.\,\ref{fig5}(b). Up to an applied magnetic field of $\mu_0H = 0.8$\,T, at RT and low temperature the resistivity change shows approximately the same magnitude. However, the low temperature (150 K and 107 K) resistivity continues to decrease with the same rate up to a magnetic field of $\mu_0H = 1$\,T reaching $\Delta \rho = -1.95$\,\% with a further moderate decrease to $\Delta \rho = -2.0$\,\% at $\mu_0H = 2$\,T. At a temperature of $T = 107$\,K, $\Delta\rho$ shows oscillations around the smoothly decreasing GMR curve measured at $T = 150$\,K. However, the temperature was no less stable in this case. We have no explanation for this effect.

The temperature dependence of the electrical resistance in zero magnetic field clearly shows a metal-like behavior (Fig.\,\ref{fig5}(c) for temperatures between 100\,K and 200\,K with an extrapolated residual resistivity of 4.3\,$\mu\Omega\cdot$cm).

\section{Results -- Influence of thermal treatment}

\subsection{Development of structural parameters observed by XRD}

\begin{figure}
  \includegraphics[width=7.5cm]{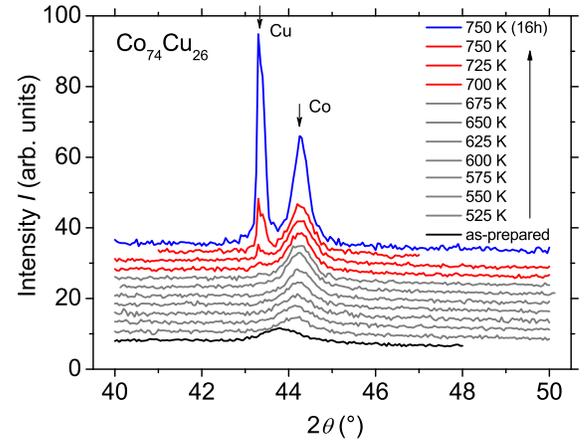}
    \caption{\label{fig6} (Color online) Development of the diffraction pattern (2$\theta$ range near the 111 Co-Cu reflection) for a Co$_{74}$Cu$_{26}$ film after annealing at the temperatures indicated. The annealing time was 75\,min except for the topmost pattern (16\,h). All the diffraction patterns were taken at RT.}
\end{figure}

\begin{figure}
  \includegraphics[width=7cm]{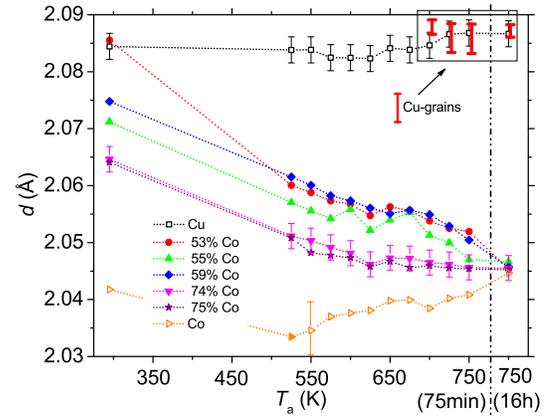}
    \caption{\label{fig7} (Color online) Development of the (111) lattice spacing of Co$_{x}$Cu$_{100-x}$ nanolayers for different Co concentrations as a function of annealing temperature $T_{a}$. The observed lattice spacings cannot be attributed to a solid solution (see discussion).
For $T_{a} > 675$\,K the used values are obtained from the right reflection (111 Co) in the WAXRD pattern (see Fig.\,\ref{fig6}) and are represented by solid symbols similar to $T_{a} < 675$\,K. For $T_{a} > 675$\,K the values determined from the left reflection (111 Cu) in the WAXRD pattern (see Fig.\,\ref{fig6}) do not change significantly and are indicated by thick (red) bars (upper right part, framed). The lines are guides to the eye. E.s.d.'s of the experimental points are shown by the vertical bars for pure Cu, Co, and Co$_{74}$Cu$_{26}$.}
\end{figure}

The development of the X-ray diffraction pattern due to thermal treatment is shown in Fig.\,\ref{fig6}. The corresponding lattice spacings are shown in Fig.\,\ref{fig7}, where the monotonic decrease of the lattice spacing $d_{111}$ of the Co$_{x}$Cu$_{100-x}$ alloy nanolayers with increasing annealing temperature for $T_a < 675$\,K and for Co-rich grains to higher temperatures is apparent (see solid symbols for alloy compositions in Fig.\,\ref{fig7}). As a result of annealing process, the interplanar distances $d_{111}$ of the Co$_{x}$Cu$_{100-x}$ alloy nanolayers show a better agreement with Vegard's law, with values slightly smaller than could be expected from the law (see Fig.\,\ref{fig2} as an example for annealing at $T_a = 525$\,K and $T_a = 650$\,K). These results show that it is also possible to adjust the (average) (111) lattice spacing of the material system Co-Cu with sub-picometre precision by using a thermal treatment. 

The development of the lattice spacing belonging to copper-rich grains is shown in the upper right part of Fig.\,\ref{fig7}, where all measured values are within the thick (red) bars. The corresponding WAXRD reflection, appearing after a thermal treatment at 425\,K (Fig.\,\ref{fig6}), does not undergo a significant change in 2$\theta$ for the last four thermal treatments but its intensity does. The lattice spacing calculated for this reflection corresponds to pure copper, where the presence of copper has been verified by selected area EDX. The two resulting reflections formed after the final thermal treatment belong to pure cobalt and pure copper. Fig.\,\ref{fig3}(c) shows the development of the crystallite size as a function of the annealing temperature for a pure copper sample and for Co$_{53}$Cu$_{47}$. Up to $T_a = 675$\,K the crystallite size shows a moderate increase with increasing temperature. This is followed by a significant growth of Cu-rich crystallites after higher temperature anneals, where the individual reflections of Co- and Cu-rich grains are formed (see Fig.\,\ref{fig6}). 

Note that after annealing at temperatures higher than 425\,K, the crystallite/grain sizes determined by WAXRD and TEM exceed the film thickness. However, this possible contradiction is resolved taking into account that only lateral grain size is accessible by TEM. In the case of WAXRD, the calculated crystallite size is a mean effective quantity averaging the crystallite sizes in different directions.  

\subsection{Thermal treatment during TEM investigations}
An additional thermal treatment test was performed in the TEM for temperatures between 450\,K and 600\,K. After 15 min anneal, a bright field image and an ED pattern were taken after the sample had cooled down to RT. 
An increase of the grain size during the thermal treatment is observable from the lower part of Fig.\,\ref{fig4}. The dark regions, with a size of about 100 nm, belong to grains of pure copper, as verified by selected area EDX. Similar values for the crystallite sizes of Cu-rich grains after annealing are obtained by XRD using the Scherrer equation (Fig.\,\ref{fig3}(c)). In the upper part of Fig.\,\ref{fig4} the corresponding ED patterns are shown. It can be seen that the additional rings, which do not belong to the \textit{fcc} lattice, and are attributed to an intermetallic phase of Cu$_3$Au structure-type (see Sec. 3.2), disappear completely during the thermal treatment. 

\subsection{Influence of thermal treatment on the GMR}

No direct correlation between the composition of the as-prepared or annealed layers and the magnitude of the GMR effect was found for the samples investigated in this paper.

\begin{figure}
  \includegraphics[width=7.5cm]{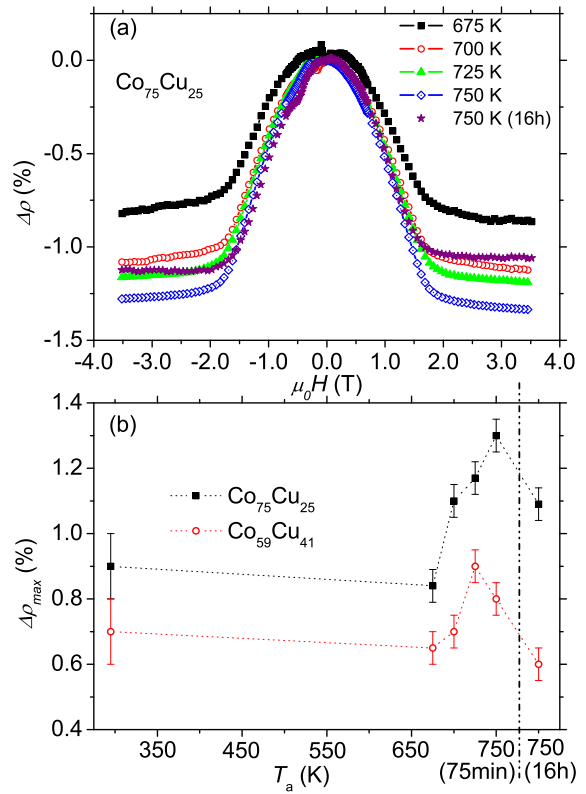}
    \caption{\label{fig8} (Color online) (a) Dependence of $\Delta\rho$ on $\mu_0H$ after annealing at given temperatures $T_{a}$. (b) Maximum $\Delta\rho_{max}\,=\,\Delta\rho(\mu_0H\,=\,3.5$\,T) as a function of annealing temperature $T_{a}$. The duration of thermal treatment was 75\,min per step except for the 16\,h for the last anneal. The lines in (b) are guides for the eye only. E.s.d.'s of the experimental points are shown in (b) by the vertical bars. The resistivity measurements in (a) and (b) were performed at RT.}
\end{figure}

The change of the GMR effect after a thermal treatment is shown in Fig.\,\ref{fig8}(a) for a nanolayer with composition Co$_{75}$Cu$_{25}$. The effect increases for anneals of up to 750\,K for 75 min. Further annealing at 750\,K for 16\,h leads to a decrease of the effect. Fig.\,\ref{fig8}(b) shows the maximum change of the electrical resistivity as a function of the annealing temperature for two samples with characteristic compositions of 59\,\% Co and 75\,\% Co, and a film thickness of about 20\,nm. Both samples show a maximum in the GMR effect after annealing at $\approx725$\,K. The magnitude of the GMR effect after a thermal treatment is a factor of $\approx 1.5$ larger than the as-prepared samples. This change is rather weak compared to other results published which show comparable values for the as-prepared samples \cite{errah00,jackson96}. However, the layer thickness was $\approx 100$\,nm  and $200$\,nm \cite{errah00, jackson96}, and annealing led to an enhancement of the GMR by a factor of $\approx 8$ and $\approx 4$, respectively. This suggests the grain size perpendicular to the electrical current plays an important role. A second possible reason for the comparatively small GMR in our case is the large cobalt content well above the percolation threshold. 

\section{Discussion}

All investigated as-prepared Co-Cu layers exhibit only one significant Bragg reflection and possess a GMR effect at the same time.
Similar observations were published several times, see e.g. \cite{errah00}. These two points are contradictory because ferromagnetic grains separated by a non-magnetic spacer, i.e. two separate phases, are necessary for the occurrence of the GMR effect. 
However, for this case, at least two Bragg reflections of two different crystalline phases, namely one for Co and one for Cu, would be expected in the WAXRD pattern, which was not the case. Furthermore, the development of the diffraction pattern after the thermal treatments (Fig.\,\ref{fig6}) could be misinterpreted as a decomposition of an initial solid solution into grains of pure Co and pure Cu. 

These discrepancies can be solved by taking into account that the grain size of $\approx10$\,nm is smaller than the coherence length $l$ of the used X-ray radiation. The lateral coherence length $l$ of the X-ray radiation can be estimated using the equation \cite{attwood} $l = s\cdot\lambda/(2\pi B$), where $\lambda = 1.54056$\,\AA\, is the wavelength used, $s = 0.22$\,m is the distance between the X-ray source and sample, and $B = 0.04$\,mm is the X-ray focus spot size (Bruker KFL-Cu-2K tube). From the equation $l\approx135$\,nm is calculated. Therefore, a coherent superposition of X-ray waves scattered by Co grains and Cu grains, results in only one Bragg reflection in the diffraction pattern (i.e. the scattered amplitudes of Co and Cu are added but not their squares \cite{meyerpaufler00}). This leads to the misinterpretation of the existence of a solid solution (mixed crystal) in the Co-Cu layer. During the thermal treatment, the increase of grain size leads to the development of two Bragg reflections of Co- and Cu-rich phases in the WAXRD pattern as soon as the grain size exceeds the coherence length of the radiation. This is supported by TEM results which show copper grains larger than 100\,nm (lower panel in Fig.\,\ref{fig4}(b)). Furthermore, the size of the copper crystallites is estimated to be larger than 100\,nm using the Scherrer equation (Fig.\,\ref{fig3}(c)) with the FWHM of the developing 2nd Bragg reflection attributed to pure Cu (Fig.\,\ref{fig6}, $T_a > 700$\,K, left WAXRD reflection). 
Therefore, the change of lattice spacings shown in Fig.\,\ref{fig7} is not an intrinsic property of the Co-Cu layers but is caused by a complex change of the interference conditions due to increasing grain size. As a matter of fact, the lattice spacing of Co- and Cu-grains in the alloy films does not change significantly during the annealing process, similar to the lattice spacing of single phase Co- and Cu-films.

\begin{figure}
  \includegraphics[width=7cm]{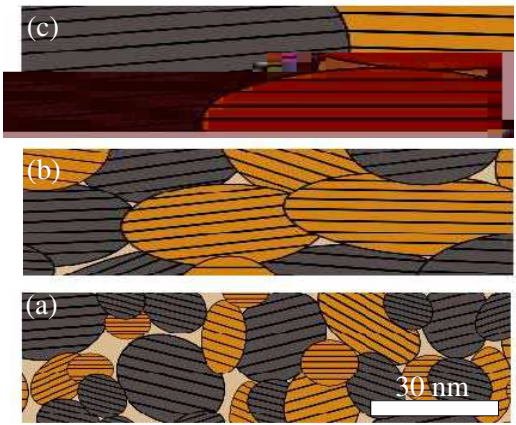}
    \caption{\label{fig9} (Color online) Schematic plot of real structure and its development due to a thermal treatment for a granular Co-Cu nanolayer (Co and Cu grains are shown by dark gray and light orange colors, respectively). 
    (a) as-prepared film with fine grains of Co and Cu with mean size $D$ less than the coherence length $l$ of the X-ray radiation, (b) film after annealing at $T = 700$\,K with Co and Cu grain mean size $D$ of about $l$, and (c) film after annealing at $T = 750$\,K with Co and Cu grain mean size $D$ larger than $l$.}
\end{figure}

A model of the real structure of Co-Cu nanolayers and their development after thermal treatment is schematically represented in Fig.\,\ref{fig9}. The as-prepared film material consists of fine Co- and Cu-grains which are smaller than the coherence length of the X-ray radiation as supported by TEM and WAXRD measurements. As a result, the WAXRD patterns exhibit only one Bragg reflection, the electrical resistivity measurements reveal the presence of the GMR effect. Thermal treatment leads to an increase of the grain size and enhanced GMR values. When the (lateral) grain size reaches the coherence length of the X-ray radiation (Fig.\,\ref{fig9}(b)), a second Bragg reflection is observed which corresponds to the lattice spacing of pure copper. Annealing at higher temperatures results in a further increase of (lateral) grain size (Fig.\,\ref{fig9}(c)) and an increase in GMR effect. However, prolonged annealing at high temperatures results in a decrease of GMR (Fig.\,\ref{fig8}(b)), which could be caused by an enhanced spin-flip scattering inside larger crystallites \cite{mathon01}. Finally, for higher annealing temperatures, two distinct Bragg reflections occur in WAXRD patterns and large Cu grains become visible in the TEM bright field images.

\section{Conclusions}

The as-prepared film material consists of fine Co- and Cu-grains with typical sizes smaller than the coherence length of the X-ray radiation (Cu-$K_{\alpha}$) used. 
This gives an explanation for the occurrence of the GMR effect in as-prepared and annealed films, even though X-ray diffraction patterns show only one Bragg reflection.
The size of crystallites in films is more strongly affected by the Co content than by the layer thickness.
No intermetallic alloy was found in as-prepared and annealed Co-Cu films on silicon substrates using WAXRD. 
However, indications for an intermetallic alloy of Cu$_3$Au structure-type in as-prepared Co-Cu films were observed in ED-patterns, showing that this phase exists, as theoretically predicted in \cite{yan07}. 
Annealing of the samples results in the growth of larger Co- and Cu-rich grains and a larger GMR effect.
However, the appearance of larger GMR values is assumed to be restricted by the small grain size perpendicular to the electrical current, i.e. by the layer thickness.

\section{acknowledgment}
The authors would like to thank M. D\"orr and S. Granovsky for help with electrical transport measurements and R. Boucher for valuable discussions.

\end{document}